\documentstyle[epsf,11pt]{article}
\def\baselinestretch{1.3}
\title{\bf Statistically Significant Strings are Related to Regulatory Elements in the Promoter Regions of {\it Saccharomyces cerevisiae}}
\author{Rui Hu$^{1,2}$, Bin Wang$^1$\\
  {\small $^1$Institute of Theoretical Physics, Academia Sinica}\\ 
   {\small P. O. Box 2735, Beijing 10080, China}\\ 
   {\small $^2$Department of Mordern Physics,} \\
    {\small  University of Science and Technology of China,}\\
     {\small Anhui, 230027, China}
    }
\date{}

\begin{document}
\maketitle
\begin{abstract}

Finding out statistically significant words 
in DNA and protein sequences forms the basis for many genetic studies. 
By applying 
the maximal entropy principle, we give one systematic way to 
study the nonrandom occurrence of words in DNA or protein sequences.
Through comparison with experimental results, it was shown that patterns of regulatory binding sites in {\it Saccharomyces cerevisiae}({\em yeast}) genomes tend to occur significantly in the promoter regions. We studied two correlated gene family of {\em yeast}. The method successfully extracts the binding sites varified by experiments in each family. Many putative regulatory sites in the upstream regions are proposed. The study also suggested that some regulatory sites are active in both directions, while others show directional preference. 

\end{abstract}

\section{\bf Introduction.} 

It is attractive, but not unexpected, that DNA and protein sequences deviate 
remarkably from random sequences~\cite{r1}.
According to information theory, random sequences carry minimal 
information (maximal entropy)~\cite{r2}, while the total information of life is assumed to be in DNA and protein sequences. 
As a result, investigation on 
the non-randomness of DNA and amino acid sequences 
would be the focus of Bioinformatics.

To find out nonrandom occurrence of words (short strings) in 
non-coding DNA sequences 
is interesting because a large portion of regulatory elements of eukaryotes 
usually are words of limit length in the non-coding sequences 
(for example, about 10 bases,
while the core part is about 5 bases~\cite{b1}). subjected to 
functional constraints, the patterns of regulatory elements are expected to deviate from random occurrence. 

In this paper, by applying the Maximal Entropy Principle (MEP), we develop one way to investigate the nonrandom occurrence of words in DNA sequences. Each word is given one significance index which quantifies the nonrandomness occurrence of the word. The method is then applied to study the promoter regions of {\it Saccharomyces cerevisiae} ({\it yeast}).~\cite{ar1} 
We compare the theoretical result with experiments in two ways. In the first way, the promoter database of {\em yeast} (SCPD)~\cite{r10} was analysed. It was found that, statistically, overrepresented words are more easily encountered in the database. The second way is to study the promoters of coregulated gene families. The experimentally found binding sites were successfully extracted, and more putative binding sites are suggested.  

In the following the method will be developed in details, and in the third section the method will be applied to study the promoter regions of {\em yeast}.

\section {\bf Treat the nonrandomness of DNA sequences via Maximal Entropy Principle.} 

The idea comes from a simple observation. Take a long DNA sequence as 
an example. Given only the (normalized) frequencies of $A$,$C$,$G$,$T$ ($P_A$,$P_C$,
$P_G$,$P_T$) , one would expected that the frequencies 2-tuples have the form 
\begin{equation} \label{eq:intu}
	{P^0}_{c_1c_2}=P_{c_1}{\times}P_{c_2}
\end{equation}
Here $c_1$ and $c_2$ are one of the four bases ${A,C,G,T}$.

Comparison between the measured frequency $P_{c_1c_2}$ and the expected value ${P^0}_{c_1c_2}$ reveals the statistical significance of $c_1c_2$ in the sequence.

To generalize the above idea, one encounters the problem to predict the 
frequencies of $k+1$-tuples from the frequencies of $k$-tuples when $k>1$. 
A reasonable defination can then be used to evaluate the statistical significance of words longer than two bases.

The following is an attemption to answer this problem.
In the treatment, when the composition of a $k$-tuple 
is concerned, the word
will be written as $c_1c_2{\cdots}c_{k-1}c_{k}$. However when only the length 
$k$ of the word is relevant, it will be given in the form of $w^k$. 
 A combinatory form may also be used.  For example,
$w^kc$ ($cw^k$) is the word obtained by adding a letter $c$ to the 
right (left) of $w^k$. The measured and expected frequencies of $w^k$ in the
sequence will be written as $P_{w^k}$ and ${P^0}_{w^k}$, respectively.

There are a total of $4^k$ $k$-tuples. 
For prediction the Maximal Entropy Principle (MEP) is a prefered choice.
According to modern genetics, the driving force for 
nucleotide sequence evolution is, on one hand, random mutations of bases that
 maximize the entropy, and, on the other hand, the natural selection
 which subjects the maximization of entropy to certain constraints. Therefore, DNA sequence analysis shows intrinsic correlation 
to the MEP.
One brief introduction (which is necessary for our use) to MEP will be given 
below. More details can be found in e.g.~\cite{r7}.

Suppose that $\{P_i, i=0,1,2,{\cdots}\}$ is a discrete distribution. An information 
entropy can be defined on it~\cite{r2}:
\begin{equation} \label{eq:entropy}
	S=\sum_iP_ilnP_i.
\end{equation}
Usually \{$P_i$\} satisfies some constraints: 
\begin{equation} \label{eq:constraints}
F_j(\{P_i\})=0,  \qquad      	j=1,2,{\dots},M.
\end{equation}
Here $M$ is the number of constraints. Define a target function:
\begin{equation}
	H=S+\sum_{j=1}^M\lambda_j F_j(\{P_i\}),
\end{equation}
$\lambda_j$ being Largrange factors. MEP states that the distribution 
minimizing the target function $H$ is the most reasonable
distribution satisfying constraints (\ref{eq:constraints}). 
This, however, does not state 
that \{$P_i$\} is the only distribution satisfying (\ref{eq:constraints}).

The MEP now can be applied to study the problem raised above. 
The entropy function here is:
\begin{displaymath}
S=\sum_i{P^0}_{w^{k+1}(i)}ln{P^0}_{w^{k+1}(i)},
\end{displaymath}
where $i$ is a index used to distinguish k-tuples from each other. 
(In order to get the index of 
a word, the following maps were used: $A$ 
to 0, $C$ to 1, $G$ to 2, and $T$ to 3. The original word is thus 
mapped to a string containing only 0,1,2 and 3. The string is then 
considered as quaternary 
number. After being transformed to decimal, the number is used as the 
index of the word.)

Constraints in the present problem is:
\begin{eqnarray} \label{eq:normal}
P_{w^{k}(i)}=\sum_{c}{P^0}_{w^{k}(i)c},\nonumber \\
P_{w^{k}(i)}=\sum_{c}{P^0}_{cw^{k}(i)},\nonumber \\
i=0,1,2,{\cdots},4^{k}-1.
\end{eqnarray}
${P^0}_{w^{k+1}}$ is the frequency needs to be predicted and $P_{w^k}$ is the frequency already known.
There is a total of $2{\times}4^{k}$ constraints. It is possible that these 
constraints are linearly related, so that the number of effective 
constraints is smaller than $2{\times}4^{k}$. This, however, does not alternate
the result.

The solution can be obtained:
\begin{equation} \label{eq:mep}
{P^0}_{c_1c_2{\cdots}c_{k+1}}=\frac{P_{c_1c_2{\cdots}c_k}{\times}P_{c_2c_3{\cdots}c_{k+1}}}{P_{c_2c_3{\cdots}c_{k}}}.
\end{equation}
When k=1, the solution reduces to the intuitively result, eq.(\ref{eq:intu}).

The above treatment is from $k$-tuples to $k+1$-tuples. As a generic 
scheme, the MEP can also be applied to predict the frequencies of $k+2$-tuples, 
$k+3$-tuples and so on,  
based on the frequency of k-tuples.
Actually, one can get the result by repeatedly applying eq. (6). For example:
{
\setlength\arraycolsep{2pt}
\begin{eqnarray} \label{eq:mlevel}
{P^0}_{c_1c_2{\cdots}c_{k+1}c_{k+2}}&=&\frac{{P^0}_{c_1c_2{\cdots}c_{k+1}}{\times}{P^0}_{c_2c_3{\cdots}c_{k+2}}}{P_{c_2c_3{\cdots}c_{k+1}}} 
	\nonumber \\
&=&\frac{P_{c_1c_2{\cdots}c_k}{\times}{P_{c_2c_3{\cdots}c_{k+1}}}{\times}P_{c_3c_4{\cdots}c_{k+2}}}{P_{c_2c_3{\cdots}c_k}{\times}P_{c_3c_4{\cdots}c_{k+1}}}.
\end{eqnarray}
}
Thus, when one refers to the expected frequency of a certain word of length $k$, 
the knowledge that the prediction is based on must be pointed out. 

With the frequencies of longer words, 
one can always obtain the frequencies of shorter ones. 
On the other hand, the expected frequencies of longer words, 
eq.(\ref{eq:mep}), is predicted from the frequencies of shorter words, with no more 
information added. Therefore,
the deviation of the measured frequencies from the expected ones gives new 
information emerges only in the frequencies of the longer words. 
In order to use this part of information, we refer to the following significance index 
\begin{equation} \label{eq:index}
I_{w^k}=\frac{P_{w^k}-{P^0}_{w^k}}{\sqrt{{P^0}_{w^k}}}.
\end{equation}
The indexs of $k$-tuples form a vector of $4^k$ dimension. 

It should be pointed out that the simple solution eq.(\ref{eq:mep}) 
results from the constraints, eq.(\ref{eq:normal}). Although there are many ways to write down the prediction~\cite{c1,b2}, the Maximal Entropy Principle ensure that, 
submitted to these constraints, the solution eq.(\ref{eq:mep}) is the best 
one. 
However, one can consider more constraints. Expect for the continuous words, spaced patterns can also be involved in the above statistical treatment~\cite{b2}. As an example, consider the spaced word $c_1$-$c_2$, where $c_1$ and $c_2$ are certain bases and the base between them is not relevant. One more constraint 
\begin{displaymath}
	P_{c_1-c_2}=\sum_{c}P_{c_1cc_2}
\end{displaymath}
can be added to the frequencies of $3$-tuples, and the statistical significance of the spaced words can also be evaluated. The MEP, as a general framework, is still applicable, but there will be no simple explicit solution as eq.(\ref{eq:mep}).

\section{\bf The relationship between regulatory elements and statistically 
significant words in the {\em yeast} promoter regions.}

With the accumulation of huge amount of genome sequences, analysis of
the regulatory regions becomes urgent, because they govern the regulation 
of gene expression. Finding out the regulatory sites in Eukaryotes genomes is 
especially difficult, largely because of their strong variance. 
This, however, gives the chance for 
statistical methods to play an important role in binding sites prediction. 

The regulatory elements are functionally constrained and are often
 shared by many genes. As a result, the sites are expected to be 
significantly represented. Based on this 
belief, the method developed above is expected to be applicable 
in finding regulatory sites in the promoter regions of {\it yeast}.
we employ two ways to check this point.

In the first way, as just an illustration of 
the effectiveness of the MEP treatment, a data set including
all the promoters of {\it yeast} will be used 
to perform the statistical 
evaluation. The promoter regions refer, 
according to Zhang~\cite{b1}, to the upstream region of 500 bases long. 
From the sequence set the word frequencies are obtained 
 and ${I_{w^k}}, k=2,\cdots,8$, are calculated according to eq.(\ref{eq:mep}) and eq.(\ref{eq:index}).
(to obtain ${I_{w^k}}$, ${P^0}_{w^k}$ is predicted based on the frequency of k-1-tuples.)
For comparison the index ${I_{w^k}}, k=2,\cdots,8$, of words in the coding regions (CDSs) of {\em yeast} were also calculated. 

To compare the significance index of words with experimentally verified regulatory 
 elements, a strongly statistically characterized method was pursued. 
The promoter database of {\it yeast} collected by 
Zhou et al.~\cite{r10} was used as targets. 
One word is called to {\it hit the target} if it covers a known 
regulatory element or part of the element. 
In this way, each word will be checked against all the elements in 
the database. We want to see if the total hits of words show correlation 
with the significance index. 

Fig.1 shows the ratio of the average 
hits of words whose significance index are larger than 
a certain cutoff (5.0, 3.0, or 2.0) to the average hits of all the $k$-tuples. 
Some properties of significance index in the promoter regions are revealed.
First, for all the cutoff value shown in fig.1, the ratios are always larger than $1$.
Second, when the words are longer than 4 bases, the average hits 
increase with the increase of cutoff. Furthermore, the ratio also increases with the increase of word length.
As a comparison, Fig.1 shows that the ratio of hits does not depend on significance index in the CDS regions.

To see the dependence of hits on significance index further, words are divided
 into groups according to their significance index values. In each group the hits were 
averaged. See table 1, and Fig.2 which is based upon the
data in Table.1 but shown as a more audio-visual illustration. The dependence of hits on significance index shown in Fig.1 is seen again. 
Furthermore, the average hits are not the 
monotonic function of significance index in the promoter regions. 
For words with both positively and 
negatively large significance index in the promoter regions, 
the average hits are larger than those of words 
whose significance index is around zero. 
Again no dependence of average hits on significance index 
in CDS regions is observed in Table.1 and Fig.2.

That words with large {\em negative} significant index 
in the promoter regions also show 
higher affinity to binding sites deserves more consideration. One account is that although some regulatory elements, such as those involved in the expression of housekeeping genes, are expected to be overrepresented since large amount of the genes are needed, others that control the expression of some essential but restrictedly needed genes, are expected to be underrepresented to avoid inappropriate translation. However, more convincing explaination exists: if a word, e.g., $w$A, has high positive index, then some of $w$C,$w$G,$w$T are expected to have negative index. This can be seen from the following example. While the index of TATAT is 16.3, that of TATAA is -12.2. Actually, both have much high counts in the sequences and both are variance of binding site of the same transcriptional factor. 

For universally existing regulatory elements, as expected, the 
significance index in 
the promoter regions are much high. One example is the poly(A/T) stretches. 
As given above, the significance index of TATAT is 16.3. Also the significance 
index of TATATAT, 8.1, is high.
As another example, the significance index of the core of CAAT-box, CAAT, 
is 8.95. 
However, in order to develop an algorithm for regulatory elements prediction, 
more subtle consideration must be involved. 
First, genes are needed to be classified into 
families to improve the compositional bias of the sequences. 
Furthermore, more complicated usage of the information given 
by significance index should 
be considered, because, according to eq.(\ref{eq:mlevel}), the expected frequency of $k$-mers can be defined in $k-1$ ways, i.e., based on the frequency of $1,2,\cdots,k-1$-mers, respectively. For each definition the significance index can be obtained. 
On considering the statistical property of words in the sequences, each of these indexes would give useful information.
We choose two coregulated gene family to further test our method.

The coregulated genes of {\em yeast} metabolism have been widely studied, and these datasets provide ideal material to test the methods for binding sites prediction. Two families of coregulated genes, GCN and TUP, were shown in table 2. Detailed information on them can be found in~\cite{r9}. For each family, the frequencies of 6-tuples in the promoter regions were first collected. The expected frequencies them were predicted in five ways, which are based on the frequencies of bases, 2-tuples, 3-tuples, 4-tuples, 5-tuples, respectively. In stead of $I_{w^k}$, a simpler significance index $P_{w^k}/{P^0}_{w^k}$ was used. In our study only the single strand of promoter sequences is considered. This is different from that of~\cite{r9}. They count the number of each words in both strands. In this way there are only 2080 distinct oligonucleotides, while the number in ours is $4^6=4096$. Table 3 shows the words that possess no less than 3 among the 5 significance index larger than 3.  There are 13 such words for GCN family, and 23 for TUP family. In table 3 several words tend to cluster together to form a longer pattern. Generally speaking, the clusters can be expanded by involving  words with slightly lower significance.  

In both families, 6-tuples corresponding to regulatory binding sites found by experimental analysis are observed in table 3. See the first cluster of words for GCN family and the first and the second clusters for TUP family. Most of these words also show high statistical significance in the analysis of~\cite{r9}. Some words predicted by~\cite{r9} but not varified by experiments are also observed in table 3 (significant words shared by~\cite{r9} and the present analysis are shown in bold in table 3). However, our analysis also found many significant words which do not show as highly significant scores according to~\cite{r9}.
 
Two clusters of words for TUP families is noteworthy (see the first and the second clusters in table 3). The first cluster includes GTGGGG, AGGGGC, ACGGGC, TGGGGT, and GGGGTA, and the second cluster involves TACCCC, ACCCCG, CCCCGC, and CCCCAC. between them GTGGGG and CCCCAC, GGGGTA and TACCCC are reverse complements. The two clusters both correspond to the binding sites of transcription factor Mig1p (Zn finger), but seen from different strands. This may imply that the binding sites of Mig1p are active in both orientation. This property, however, was not found for the binding sites of Gcn4p (see table 3). For example, when the cutoff of significance index is reduced to 1.3 (now 46 words satisfy the creterion), the cluster of TGACTC and GACTCA expands to involve another 4 members: CGATGA, GATGAC, ATGACT, and GTGACT; while only one of inverse complements of them, GAGTCA, also has 3 index larger than 1.3. Among the 46 words, it can only be clustered with another words GGAGTC. Thus, the binding sites of Gcn4p seem to be active preferencially in one direction.

Among the available methods of binding sites prediction, ours is similar to that of~\cite{r9} in that both work by defining expected frequencies of words. the difference is that our method defines the expected frequences on the statistical stproperties of the sequences themselves, while~\cite{r9} more or less heuristically defines the word frequencies of whole non-coding sequences as the expected value. It is thus expected that our method is more precise and gives more unbiased result. 

An alternative method developed by Li et al~\cite{add4}. gives more subtle consideration on the statistical feature of DNA sequences. In their model, the sequence is considered as a text without interwords dilimiters. They apply maximal likelihood consideration to {\em recover} the words, which they consider as possible binding site condidates. But the computation is far more complex to get meaningful result. 

More methods to detect unknown elements within funtionally related sequences are availible (for a review, see \cite{add1}), most of which, such as the consensus~\cite{add3} and the Gibbs sampler~\cite{add2}, are based upon well difined biological models. The type of signals that can be detected are generally limited; it is difficult for them to detect multiple signals. But these methods are able to detect much larger patterns with high precision. The present method can be used to detect multiple elements, but the pattern it can find is short. 

It is also a widely explored problem in biology to compare the noncoding 
and coding regions of DNA sequences~\cite{d1,d2,r11}. The MEP treatment 
 gives one systematic way to study the statistical differences between
 coding and noncoding regions.  In table.1 it is shown 
that significance index in CDS 
regions distribute much more stretchy than that 
of the promoter regions. The contrast keeps for all the word lengths we 
studied (up to 8 bases). This reveals that CDS regions are in a more 
nonrandom state. Two factors may help to interpret this phenomenon.
First, the mutation rate of CDS regions is much lower than 
that of the promoter regions~\cite{d2}. Secondly, the code usage in 
CDS region is universal and definite, while in the promoter regions 
the length of regulatory elements differ from each other and the regulatory 
elements may differ strongly from the consensus sequences. 

\section*{ACKNOWLEDGMENTS}
We are grateful to professor Bai-lin Hao and Wei-mou Zheng for stimulating discussions. We also thank Guo-yi Chen for helps on computation.

\newpage

\begin{figure}
\centerline{\epsfxsize=10cm \epsfbox{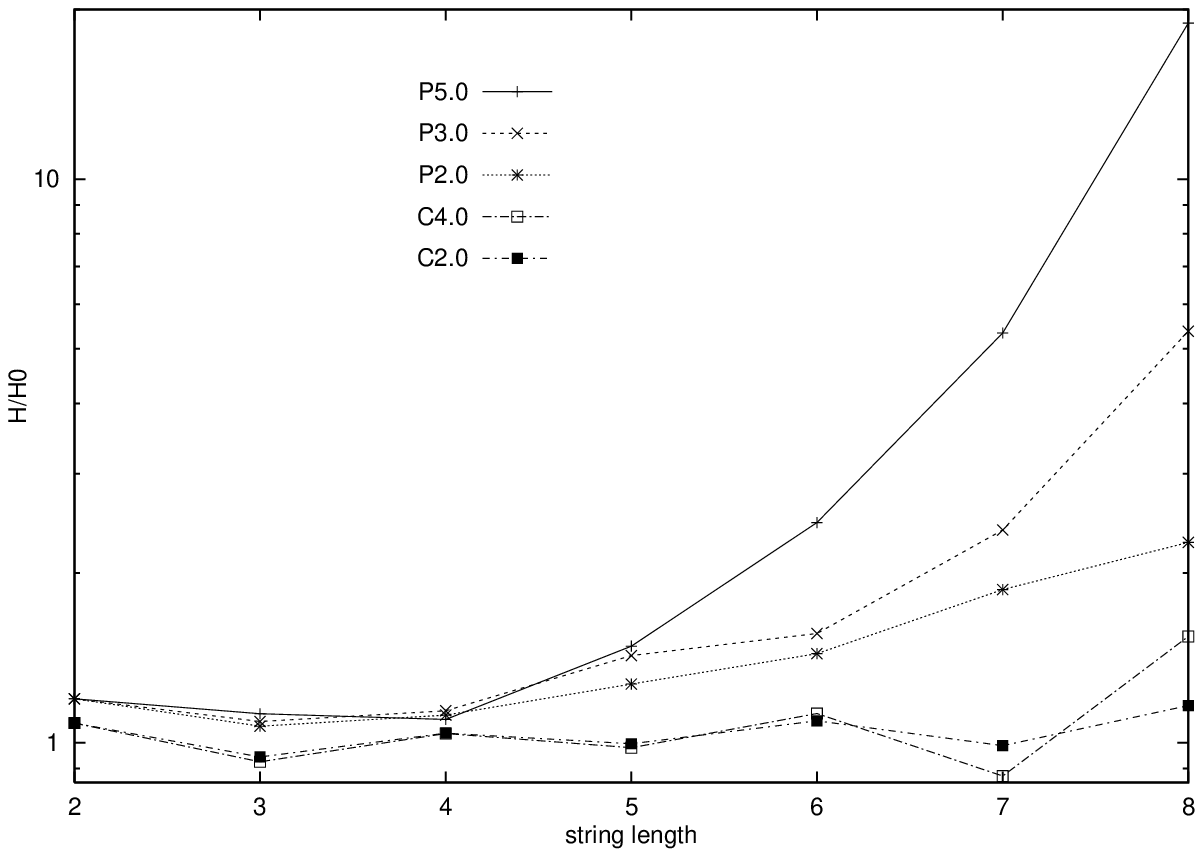}}
\caption{
The ratio of average hits ($H$) of words above certain cutoff of significance index to the average hits ($H_0$) of all the words of same length. The $H_0$(word length) are $405(2), 94.4(3), 21.7(4), 4.92(5), 1.10(6), 0.241(7), 0.0528(8)$.
}
\end{figure}

\newpage

\begin{table}
\caption{The dependence of average hits on the significance index 
$I_{w}=\frac{P_w-{P^0}_w}{\sqrt{{P^0}_w}}.$
 The values shown in the hits volume are averaged over the hits of the points (words) included in the significance index range shown in the $I_w$ colummn.}
{\scriptsize
\renewcommand{\baselinestretch}{-0.2}
\begin{center}
\begin{tabular}{|lll|lll|lll|lll|}
\hline
\multicolumn{6}{|c|}{pentemer}&
\multicolumn{6}{c|}{hexmer} \\
\hline
\multicolumn{3}{|c|}{promoter}&
\multicolumn{3}{c|}{CDS}&
\multicolumn{3}{c|}{promoter}&
\multicolumn{3}{c|}{CDS} \\
\hline
$I_w$&points&hits&$I_w$&points&hits&$I_w$&points&hits&$I_w$&points&hits \\
\hline
-15,-9&9&7.33&-29,-12&16&5.50&-11,-6&10&2.10&-15,-9&12&1.17 \\
-9,-7&12&4.75&-12,-10&12&4.33&-6,-4&15&1.40&-9,-7&15&0.60 \\
-7,-5&16&4.00&-10,-9&12&5.67&-4,-3&28&1.79&-7,-6&30&1.40 \\
-5,-4&22&3.50&-9,-8&14&5,14&-3,-2&161&1.04&-6,-5&95&0.78 \\
-4,-3&29&4.31&-8,-7&23&4.87&-2,-1&716&0.976&-5,-4&102&0.98 \\
-3,-2&91&4.02&-7,-6&40&5.38&-1,0&1137&0.997&-4,-3&202&1.15  \\
-2,-1&158&4.03&-6,-5&44&4.34&0,1&1169&1.05&-3,-2&334&1.10  \\
-1,0&182&4.46&-5,-4&46&5.41&1,2&593&1.27&-2,-1&563&1.07  \\
0,1&198&5.02&-4,-3&63&5.27&2,3&178&1.51&-1,0&738&1.09 \\
1,2&134&5.58&-3,-2&74&5.34&3,4&52&1.28&0,1&750&1.04 \\
2,3&77&5.25&-2,-1&79&4.52&4,5&19&1.68&1,2&570&1.09 \\
3,4&47&6.55&-1,0&94&5.53&5,6&11&2.18&2,3&345&1.24  \\
4,6&21&7.09&0,1&101&4.58&6,13&12&3.17&3,4&117&1.07 \\
6,8&13&6.15&1,2&83&4.16& & & &4,5&94&1.29  \\
8,10&10&7.40&2,3&59&5.47& & & &5,6&64&1.17 \\
10,19&9&10.22&3,4&60&4.60& & & &6,7&21&1.23  \\
 & & &4,5&48&4.70& & & &7,9&18&1.00  \\
 & & &5,6&38&4.08& & & &9,19&16&1.44 \\
 & & &6,7&27&4.19& & & & & &  \\
 & & &7,8&25&4.80& & & & & &  \\
 & & &8,9&19&6.84& & & & & & \\
 & & &9,10&13&5.46& & & & & &  \\
 & & &10,12&12&4.42& & & & & &  \\
 & & &12,29&19&5.21& & & & & &  \\
\hline
 \end{tabular}
\end{center}
 }
\end{table}

\newpage

\begin{figure}
\centerline{\epsfxsize=10cm \epsfbox{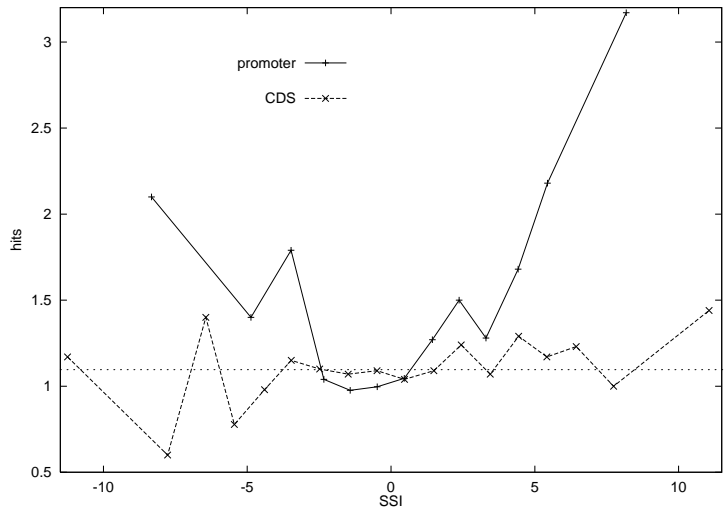}}
\caption{
The dependence of average hits of 6-tuples on their average significance index. The data in this figure are  shown as a more audio-visual illustration of the 6-tuple data in Table 1.
}
\end{figure}


\begin{table}
\caption{The coregulated gene family GCN and TUP, and criterion for them being 
clustered.}
\vspace {1cm}
{\scriptsize
\begin{tabular}{lp{5.5cm}p{4.7cm}l}
\hline
Family&Genes&Shared regulatory property&References \\
\hline
GCN&ARG1,ARG3,ARG4,ARG8,ARO3,ARO4,
ARO7,CPA1,CPA2,GLN1,HIS1,HIS2,
HIS3,HIS4,HIS5,HOM2,HOM3,HOM6,
ILV1,ILV2,ILV5,LEU1,LEU2,LEU3,
LEU4,LYS1,LYS2,LYS5,LYS9,MES1,
MET14,MET3,MET6,TRP2,TRP3,
TRP4,TRP5,THR1&General amino acid contral; 
genes activated by Gcn4p.&Hinnebusch~\cite{add5} \\
TUP&FSP2,YNR073C,YOL157C,HXT15,SUC2,
YNR071C,YDR533C,YEL070W,RNR2,
YER067W,CWP1,YGR243W,YDR043C,
YER096W,HXT6,YLR327C,YJL171C,
YGR138C,HXT4,GSY1,YOR389W,
MAL31,YML131W,RCK1&All genes which are both 
derepressed by a facter larger than 4 
when TUP1 is deleted, and 
induced by a factor larger than during the 
diauxic shift&DeRisi et al.~\cite{add6}  \\
\hline
 \end{tabular}
}
\end{table}

\clearpage
\newpage

\begin{table}
\caption{Highly overrepresented words in promoter regions of GCN and TUP family.For each family, the 6-tuples with no less than 3 among the 5 significance index larger than 3 are indicated. The words also appear in table 2 of~\cite{r9} as significant patterns are highlighted in bold. Words are clustered according to their similarity. {\em sig(i)} is the value of $P_{w^6}/{P^0}_{w^6}$ with ${P^0}_{w^6}$ being the frequency of 6-tuple $w^6$ predicted based on the frequencied of i-tuples.}
\vspace{1cm}
{\scriptsize
\begin{tabular}{lr@{}lrrrrrrcc}
\hline
\multicolumn{9}{c}{analysis result on 6-tuples}&
\multicolumn{2}{c}{sites previously characterized}\\
Family&
\multicolumn{2}{c}{Sequences}&counts&{\em sig(1)}&{\em sig(2)}&{\em sig(3)}&{\em sig(4)}&{\em sig(5)}&Consensus&binding factors\\
\hline
\vspace {-0.2cm}
GCN&
 {\bf TGA}&{\bf CTC}   &29 &4.47 &4.61 &4.16 &2.93 &1.39&RRTGACTCTTT&Gcn4p \\
& {\bf GA}&{\bf CTCA}  &21 &3.28 &3.36 &3.25 &3.24 &1.39&&(bZip) \\
\vspace {-0.2cm}
&{\bf CCGG}&{\bf TT}   &12 &3.18 &3.38 &3.47 &2.07 &1.50&& \\
\vspace {-0.2cm}
&CCGG&GT   &6 &2.77 &3.27 &3.29 &3.01 &1.70&-&- \\
&  GG&GCGC &5 &4.02 &3.10 &2.93 &3.66 &1.68&& \\
\vspace {-0.2cm}
&CAG&CAG   &16 &4.35 &3.45 &3.12 &1.99 &1.69&-&- \\
&{\bf CAG}&{\bf CGG}   &12 &5.61 &4.95 &4.63 &2.28 &1.55&& \\

&{\bf CCG}&{\bf CTG}   &12 &4.99 &4.60 &3.51 &2.18 &1.36&-&- \\
\vspace {-0.2cm}
&CCC&CCC   &7 &3.71 &3.88 &3.15 &2.10 &1.84&& \\
&CCT&GCC   &10 &3.75 &3.15 &3.22 &1.94 &1.55&-&- \\
&GTG&CCA   &14 &3.76 &3.35 &3.06 &2.23 &1.37&& \\

&GGT&GGT   &10 &3.26 &3.73 &3.14 &2.19 &1.53&-&- \\

\hline
\vspace {-0.2cm}
TUP&
 {\bf GTGG}&{\bf GG}   &9 &6.67 &5.23 &3.76 &3.27 &1.47&& \\
\vspace {-0.2cm}
& {\bf AGG}&{\bf GGC}  &10 &6.77 &3.90 &3.65 &2.64 &1.64&KANWWWWATSYGGGGW&Mig1p\\
\vspace {-0.2cm}
& ACG&GGC  &7 &4.49 &3.57 &3.23 &2.62 &1.97&&(Zn finger) \\
\vspace {-0.2cm}
& {\bf TGG}&{\bf GGT}  &9 &4.10 &3.21 &3.14 &3.39 &1.37&& \\
&  {\bf GG}&{\bf GGTA} &10 &4.39 &4.29 &3.52 &2.89 &1.58&& \\

\vspace {-0.2cm}
&{\bf TACC}&{\bf CC}   &16 &5.67 &5.73 &4.22 &2.52 &1.32&& \\
\vspace {-0.2cm}
& {\bf ACC}&{\bf CCG}  &11 &6.34 &5.24 &5.15 &3.27 &1.39&Complement of&Mig1p \\
\vspace {-0.2cm}
&  {\bf CC}&{\bf CCGC} &8 &7.37 &4.68 &3.60 &2.31 &1.33&KANWWWWATSYGGGGW&(Zn finger) \\
&  {\bf CC}&{\bf CCAC} &12 &6.55 &5.05 &3.49 &2.27 &1.36&& \\

&{\bf AGG}&{\bf AGG}   &11 &4.66 &3.79 &3.12 &1.70 &1.44&-&- \\
& GG&TGGT  &9 &4.10 &4.27 &3.41 &2.21 &1.31&& \\

\vspace {-0.2cm}
&CTC&GAG   &8 &3.15 &4.00 &4.42 &2.22 &1.17&-&- \\
& TC&GAGG  &9 &3.75 &3.88 &4.38 &2.15 &1.73&& \\

\vspace {-0.2cm}
&{\bf GCG}&{\bf GAG}   &7 &4.74 &4.07 &3.20 &1.84 &1.35&-&- \\
& CG&GAGA  &10 &4.02 &4.17 &3.05 &1.97 &1.69&& \\

\vspace {-0.2cm}
&CTG&CTA   &10 &2.42 &3.23 &4.28 &3.21 &1.90&& \\
\vspace {-0.2cm}
&{\bf GTG}&{\bf CCT}   &17 &6.95 &6.81 &4.86 &3.34 &1.71&-&- \\
& TG&CCAC  &10 &3.74 &3.38 &3.02 &1.74 &1.51&& \\

\vspace {-0.2cm}
&GCG&CCG   &4 &4.10 &3.12 &3.13 &3.23 &2.67&& \\
\vspace {-0.2cm}
&GCA&ACG   &9 &3.43 &2.88 &3.12 &3.13 &1.37&-&- \\
&{\bf GCA}&{\bf CGG}   &8 &5.13 &4.66 &3.12 &2.58 &1.66&& \\

&CAG&TGG   &8 &3.33 &3.48 &3.01 &1.90 &1.61&-&- \\

&CGC&GAT   &7 &2.76 &3.48 &4.12 &3.68 &2.083&-&- \\
\hline
\end{tabular}
}
\end{table}
\end{document}